\documentclass[12pt,preprint,psfig,epsfig,graphics]{aastex}
\def\arcs{$''$}

\begin{document}
\title{Constraints on $z\approx10$ Galaxies from the Deepest HST NICMOS
Fields}
\author{R.J. Bouwens$^{2}$, G.D. Illingworth$^{2}$, R.I. Thompson$^{3}$, 
M. Franx$^{4}$}
\affil{1 Based on observations made with the NASA/ESA Hubble Space
Telescope, which is operated by the Association of Universities for
Research in Astronomy, Inc., under NASA contract NAS 5-26555.}
\affil{2 Astronomy Department, University of California, Santa Cruz,
CA 95064}
\affil{3 Steward Observatory, University of Arizona, Tucson, AZ 85721.}
\affil{4 Leiden Observatory, Postbus 9513, 2300 RA Leiden, Netherlands.}

\begin{abstract}
We use all available fields with deep NICMOS imaging to search for
$J_{110}$-dropouts ($H_{160,AB}\lesssim28$) at $z\approx10$.  Our
primary data set for this search were the two $J_{110}+H_{160}$ NICMOS
parallel fields taken with the Advanced Camera for Surveys (ACS)
Hubble Ultra Deep Field (HUDF).  The $5\sigma$ limiting magnitudes
were 28.6 in $J_{110}$ and 28.5 in $H_{160}$ (0.6\arcs$\,$ apertures).
Several shallower fields were also used: $J_{110}+H_{160}$ NICMOS
frames available over the Hubble Deep Field (HDF) North, the HDF South
NICMOS parallel, and the ACS HUDF (with 5$\sigma$ limiting magnitudes
in $J_{110}$ and $H_{160}$ ranging from 27.0 to 28.2).  The primary
selection criterion was $(J_{110}-H_{160})_{AB}>1.8$.  Eleven such
sources were found in all search fields using this criterion.  Eight
of these were clearly ruled out as credible $z\approx10$ sources,
either as a result of detections ($>2\sigma$) blueward of $J_{110}$ or
their colors redward of the break ($H_{160}-K\sim1.5$) (redder than
$\gtrsim$98\% of lower redshift dropouts).  The nature of the 3
remaining sources could not be determined from the data.  The number
appears consistent with the expected contamination from low-redshift
interlopers.  Analysis of the stacked images for the 3 candidates also
suggests some contamination.  Regardless of their true redshifts, the
actual number of $z\approx10$ sources must be $\leq3$.  To assess the
significance of these results, two lower redshift samples (a
$z\sim3.8$ $B$-dropout and $z\sim6$ $i$-dropout sample) were projected
to $z\sim8-12$ using a $(1+z)^{-1}$ size scaling (for fixed
luminosity).  They were added to the image frames, and the selection
repeated, giving 15.6 and 4.8 $J_{110}$-dropouts, respectively.  This
suggests that to the limit of this probe ($\approx0.3L_{z=3}^{*}$)
there has been evolution from $z\sim3.8$ and possibly from $z\sim6$.
This is consistent with the strong evolution already noted at $z\sim6$
and $z\sim7.5$ relative to $z\sim3-4$.  Even assuming that 3 sources
from this probe are at $z\approx10$, the rest-frame continuum $UV$
($\sim1500\AA$) luminosity density at $z\sim10$ (integrated down to
$0.3L_{z=3}^{*}$) is just $0.19_{-0.09}^{+0.13}\times$ that at
$z\sim3.8$ (or $0.19_{-0.10}^{+0.15}\times$ including the small effect
from cosmic variance).  However, if none of our sources is at
$z\approx10$, this ratio has a $1\sigma$ upper limit of 0.07.
\end{abstract}


\section{Introduction}
The detection of galaxies at the highest redshifts ($z\gtrsim7$)
continues to be a difficult endeavor.  Due to the redshifting of UV
light into the infrared (IR) and the well-known limitations of current
IR instruments, searches for these objects require almost prohibitive
amounts of telescope time.  Nevertheless, small amounts of deep IR
data do exist, and they can be used to set constraints on very high
redshift galaxies.  One notable example is the $z_{850}$-dropout
sample compiled by Bouwens et al.\ (2004c) which, although limited by
small numbers ($\sim5$ objects) and field-to-field variations,
provided a first detection of galaxies at redshifts beyond $z\sim7$.
In this paper, we look at the prevalence of galaxies at $z\sim8-12$ by
applying the dropout technique to the wide variety of deep F110W and
F160W-band fields that have been imaged with NICMOS (hereinafter,
refered to as $J_{110}$ and $H_{160}$, respectively).  Our principal
data set in this study will be the two deep NICMOS parallels taken
with the UDF (each has $\sim200$ orbits of data), but we will
complement this field with a variety of shallower fields possessing
similar $J_{110}$ + $H_{160}$ imaging.  These fields include the
HDF-North Thompson field (Thompson et al.\ 1999), the HDF-North
Dickinson field (Dickinson et al.\ 1999), the deep HDF-South parallel
NICMOS field (Williams et al.\ 2000), and the NICMOS footprint on the
UDF itself (Thompson et al.\ 2005).  Galaxies in the redshift range
$z\sim8-12$ are currently of great interest due to indications that
reionization of the universe may have started as soon as $z\sim17\pm5$
(Kogut et al.\ 2003) and that galaxies may have played a major role in
this process (e.g., Yan \& Windhorst 2004; Stiavelli et al.\ 2004).
We take $L_{z=3} ^{*}$ to denote the characteristic luminosity of
galaxies at $z=3$ (Steidel et al.\ 1999).  AB magnitudes are used
throughout.  We assume $(\Omega_M,\Omega_{\Lambda},h)=(0.3,0.7,0.7)$
(Bennett al.\ 2003).

\section{Observations}

\textit{(a) NICMOS parallels to the UDF.}  The two NICMOS parallels to
the UDF (taken during the first and second epochs of ACS observations)
make up our primary data set.  Each parallel consisted of $\sim$200
orbits of data, split between two pointings which overlap by
$\sim$25\% of a NIC3 pointing ($\sim3.0\times10^5\,$s were taken at
the first and $\sim8.6\times10^4\,$s at the second).  The observing
time at each position was split nearly equally between $J_{110}$ and
$H_{160}$-band observations.  While reductions of these fields are
available through STScI
(http://archive.stsci.edu/prepds/udf/udf\_hlsp.html), we used the
NICMOS procedures described in Thompson et al.\ (2005) to perform our
own reduction.  The deeper of the two pointings making up each
parallel had approximate $5\sigma$ depths of 28.6 and 28.5
(0.6\arcs$\,$aperture) in the $J_{110}$ and $H_{160}$-bands,
respectively.  The shallower pointings were some $\sim$0.6 mags less
deep.  Our reductions were drizzled onto a 0.09\arcs-pixel scale, with
approximate PSFs of 0.32\arcs$\,$ and 0.34\arcs$\,$ FWHM in the
$J_{110}$ and $H_{160}$-bands, respectively.  For optical coverage on
these fields, the $V_{606}$ and $z_{850}$-band images from Galaxy
Evolution Morphology Survey (Rix et al.\ 2004: images \#25, \#42, and
\#49) were used (Blakeslee et al.\ 2003).

\textit{(b) Shallower NICMOS Fields.}  To add area at brighter
magnitudes, we included a number of other fields with deep
$J_{110}+H_{160}$ imaging in our $J_{110}$-dropout search.  A list of
these fields is provided in Table 1, together with their approximate
$5\sigma$ limiting magnitudes ($0.6''$ aperture) and selection areas.
Other multi-wavelength data for these fields include deep $UBVI$
coverage for the HDF-N (Williams et al.\ 1996), ultra deep ACS $BViz$
coverage for the UDF (Beckwith et al.\ 2005, in preparation), NICMOS
$K_{222}$ coverage of the HDF-S NICMOS parallel (Williams et al.\
2000), and deep IRAC data in the 3.6$\mu$m and 4.5$\mu$m channels for
the GOODS North and South (Dickinson et al.\ 2005, in preparation).

\section{Analysis}

SExtractor (Bertin \& Arnouts 1996) was used to do object detection
and photometry.  Detection was performed using an aggressive $2\sigma$
detection threshold on the $H_{160}$-band images.  Our catalogs were
then cleaned of contamination from more extended background artifacts
by requiring objects to be $5\sigma$ detections in a
0.5\arcs-aperture.  We used scaled aperture magnitudes (Kron 1980:
SExtractor MAG\_AUTO) for our $H_{160}$-band total magnitudes.  Colors
were also measured with scaled apertures, but with a much smaller Kron
factor to maximize the S/N.  For images with different PSFs, colors
were only measured after the higher resolution image had been degraded
to match the broader PSF and only in a circular aperture which was at
least $2\times$ the FWHM of the broader PSF (see Bouwens et al.\ 2003
and B05 for more details).

\textit{(a) $J_{110}$-dropout selection.}  Our principal selection
criterion for our sample was a $(J_{110}-H_{160})_{AB}>1.8)$ color
cut, where the $J_{110}$-band flux was replaced by its $1\sigma$ upper
limit in cases of a $J_{110}$-band non-detection.  This criterion was
chosen to minimize contamination from all objects except $z\sim2-5$
evolved galaxies (Figure 1).  In addition, objects were required to be
non-detections ($<2\sigma$) in all optical bands as a result of
absorption from the intergalactic medium and to be blue ($\lesssim1$
mag) in all $H_{160}-K$ or $H_{160}-IRAC$ colors, since only a small
fraction ($\lesssim2$\%) of star-forming galaxies have sufficiently
red UV continuum slopes $\beta$ (i.e., $\beta\gtrsim0.5$) to produce
these colors (Adelberger \& Steidel 2000; Schiminovich et al.\ 2004).

\begin{figure}[h]
\plotone{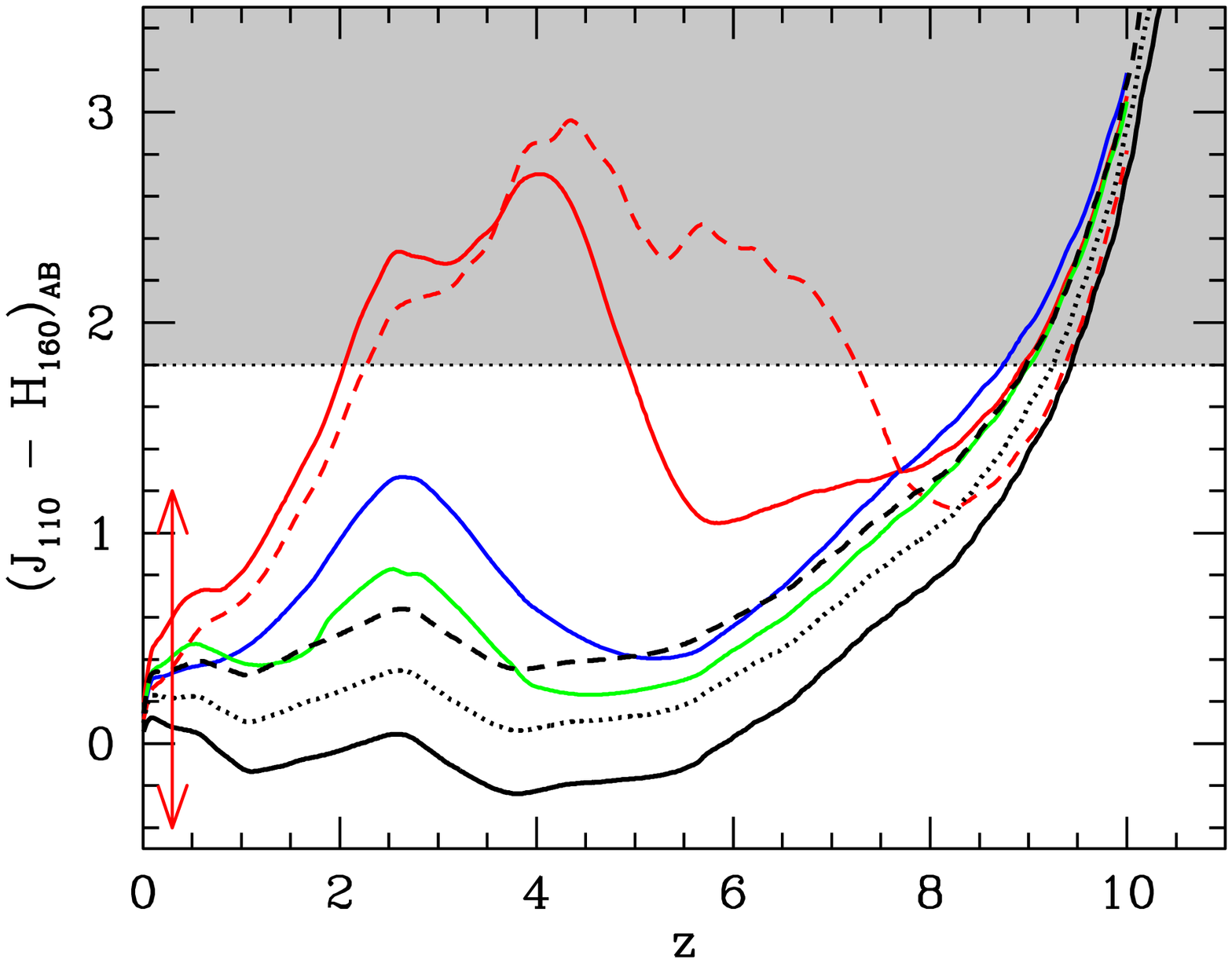}
\caption{$(J_{110}-H_{160})_{AB}$ color versus redshift for a number
of different SEDs.  Shown are the Coleman, Wu, \& Weedman (1980)
elliptical template (\textit{red line}), Sbc template (\textit{blue
line}), Scd template (\textit{green line}), and different reddenings
($E(B-V)=0.0,0.15,0.3$) applied to a $10^8$ yr burst (\textit{black
solid, dotted, and dashed lines, respectively}).  Extremely red
objects such as those found in recent IRAC selections (e.g., Yan et
al.\ 2004) have very similar colors (in the infrared) to 2.5 Gyr
bursts and are included here as the dashed red line.  The red
arrow near $z\sim0$ denotes the range of colors expected for low mass
stars (Knapp et al.\ 2004).  Our $(J_{110}-H_{160})_{AB}>1.8$
$J_{110}$-dropout criterion is shown as a dotted horizontal line and
provides a good balance between minimal contamination (selecting
evolved SEDs at intermediate redshifts $z\sim2-5$) and selecting
objects at $z\approx10$.}
\end{figure}

Applying our $(J_{110}-H_{160})_{AB}>1.8$ color cut to object catalogs
from both our primary and shallow search fields, 11 objects were found
(Table 2).  Of the 11, 6 were readily detected ($>2\sigma$) in the
bluer optical bands.  Two had very red $H_{160}-K$ colors ($\sim1.5$)
and so also appear to be low-redshift interlopers.  This was a little
uncertain for one of the two (HDFSPAR-48278437) due to the marginal
nature of its $K_{222}$-band detection, but IRAC imaging should
clarify this issue.  The final three sources (shown in Figure 2 with
two red objects from our selection) could not be excluded by either
criterion, though this may be largely due to the fact they were in the
UDF-parallel fields.  The optical imaging in these fields are not
particularly deep, nor is there any deep Spitzer or K-band data to
measure colors longward of the break.

\begin{figure}[h]
\epsscale{0.63}
\plotone{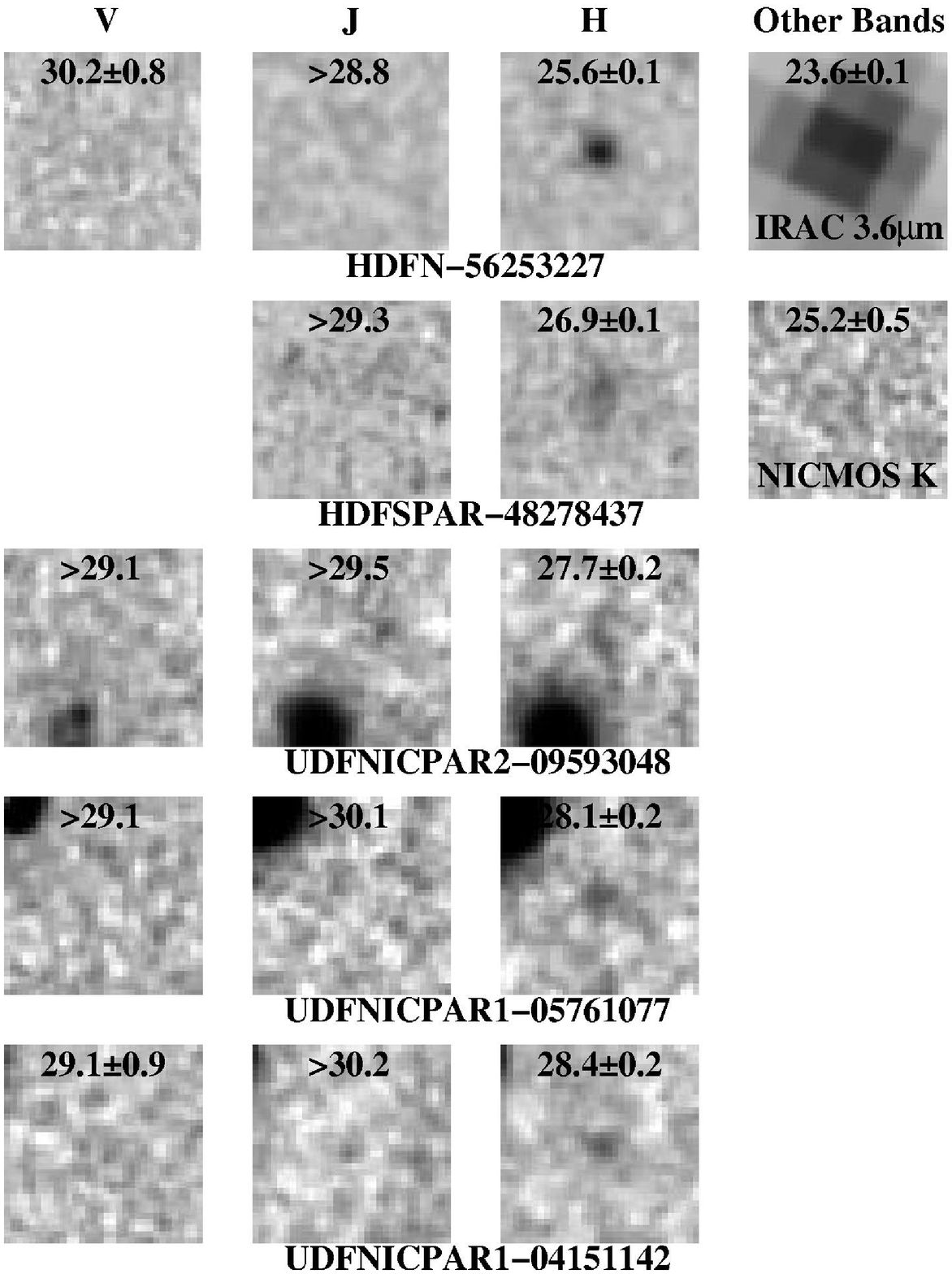}
\caption{Postage stamps images ($J_{110}H_{160}$ bands) of 5 objects
which met our $(J_{110}-H_{160})_{AB}>1.8$ selection.  The leftmost
and rightmost columns show postage images in a bluer ($V_{606}$ band)
and redder band (the IRAC 3.6$\mu$m channel for HDFN-56253227 and
NICMOS $K_{222}$ for HDFSPAR-48278437), respectively.  The overplotted
magnitudes (and $1\sigma$ upper limits) were measured within a
$0.6\arcsec$-diameter aperture.  The object in the top row
(HDFN-56253227) is the well-known Dickinson et al.\ (2000)
$J_{110}$-dropout.  Its extremely red colors ($(H_{160}-K)_{AB} = 1.5$
and $(H_{160}-IRAC_{3.6\mu m})_{AB}=2$) suggest that it is at low
redshift ($z\sim2-3$).  The second object (HDFSPAR-48278437) was
similarly excluded from our list of $z\approx10$ candidates because of
its marginal detection ($\approx2\sigma$) in the NICMOS $K_{222}$-band
and $(H_{160}-K_{220})_{AB}$ color of 1.5.  The final three objects
could be at $z\approx10$, but await limits at longer wavelengths (from
Spitzer) and good optical data to exclude the possibility that they
are highly reddened or evolved galaxies.  Objects UDFNICPAR2-09593048
and UDFNICPAR1-04151142 appear to be just detected in the
$J_{110}$-band at the $1\sigma$ level, so this may indicate they are
low redshift interlopers.  The final object (UDFNICPAR1-04151142) is
just marginally resolved, but the only stellar sources red enough to
match its colors would appear to be too bright (\S3b).  Each postage
stamp is 2.9\arcs$\times$2.9\arcs$\,$ in size.}
\end{figure}

\textit{(b) Low-Redshift Contamination.}  To help determine whether
the 3 potential $z\approx10$ candidates were low-redshift interlopers
or not, we made simple estimates of the contamination from different
types of sources.  The only known stellar sources red enough to match
these objects are extreme carbon stars or Mira variables (Whitelock et
al.\ 1995) but contamination from such sources seem unlikely due to
both their rarety and high intrinsic luminosities, which would put
them well outside the Galaxy (Dickinson et al.\ 2000).  Contamination
from redder, evolved galaxies is difficult to estimate given the large
uncertainties on the shape of the LF of these galaxies at $z\sim2-3$.
Here, we ignore these complications and simply scatter the colors of a
brighter set of galaxies ($H_{160,AB}\sim24-26$) from our fields to
fainter magnitudes.  Performing these experiments on the faint source
population from the deep parallels, we estimate $\sim1-3$ such
contaminants.  This seems consistent albeit a little less than the
$\sim3$ sources obtained.  A stack of the $V_{606}$ and $J_{110}$
exposures for all 3 sources showed detections of $0.8\sigma$ and
$1.2\sigma$, respectively, again suggesting some contamination.

\textit{(c) Expected Numbers/Incompleteness.} Having found at most 3
possible $z\approx10$ sources over our entire 14.7 arcmin$^{-2}$ search
area, it is interesting to ask how many we might have expected
assuming no-evolution from lower redshift.  Two different redshift
samples are considered as baselines: (1) a $z\sim3.8$ $B$-dropout
sample from GOODS (B05) and (2) a $z\sim6$ $i$-dropout sample from the
HUDF (Bouwens et al.\ 2005b).  As in other recent work (Bouwens et
al.\ 2004a,b,c), we project this sample to higher redshift (over the
range $z\sim7-13$: see Figure 3) using our well-established cloning
machinery (Bouwens et al.\ 1998a,b; Bouwens et al.\ 2003; B05),
accounting for pixel morphologies, the angular-size distance
relationship, NICMOS PSFs, k-corrections, and object-by-object volume
densities.  Transformed objects are added to the present NICMOS data
and the selection procedure repeated.  Direct use of the data appears
to be the best way of accounting for the substantial variations in S/N
which occur across NICMOS mosaics while including possible blending
with foreground galaxies.  Simulations were run over $\sim$30$\times$
the area of the fields.  Assuming no-evolution in size, 5.7 and 4.1
objects are expected in total (over the 6 fields) for our $z\sim3.8$
$B$ and $z\sim6$ $i$-dropout samples, respectively.  If we account for
the increase in surface brightness expected due to the
$\sim(1+z)^{-1}$ size scaling observed at $2<z<6$ (Bouwens et al.\
2004a,b; Ferguson et al.\ 2004), the expectations increase to 15.6 and
4.8, respectively.  Steeper size scalings (e.g., $(1+z)^{-2}$) yield
21.2 and 6.3 $J_{110}$-dropouts, respectively, while use of a bluer UV
continuum slope $\beta$ (e.g., $\beta\sim-2.4$) resulted in 16.2 and
4.0, respectively (assuming a $(1+z)^{-1}$ size scaling).  We note
that the two NICMOS parallels to the UDF account for $\sim73$\% of the
expected numbers (and are therefore our primary constraints), though
each of the 6 fields contributes at least a few percent.  We take the
$(1+z)^{-1}$ scaling estimates, 15.6 from $z\sim3.8$ and 4.8 from
$z\sim6$ as the most likely expected values.

\textit{(d) Previous Work.}  One previous study of $z\gtrsim8$
candidates was carried out by Yahata et al.\ (2001) on the 0.8
arcmin$^2$ NIC3 parallel to the HDF South featured here.  In that
paper, 8 $z\gtrsim10$ candidates were reported (5 of which were
selected in the $H_{160}$-band) based upon 9-band (UBVRIJHK+STIS)
photometric redshifts.  What became of these 5 candidates here?
Looking through our catalogs for this field, we found that one
(SB-NI-0915-0620) had $J_{110}-H_{160}$ colors ($\sim0.4$) that were
clearly inconsistent with a high redshift identification.  The other 4
appeared to be too faint ($H_{160,AB}\gtrsim27.5$) to set strong lower
limits on the $(J_{110}-H_{160})_{AB}$ colors and thus make robust
statements about their redshifts.

\begin{figure}
\epsscale{0.75}
\plotone{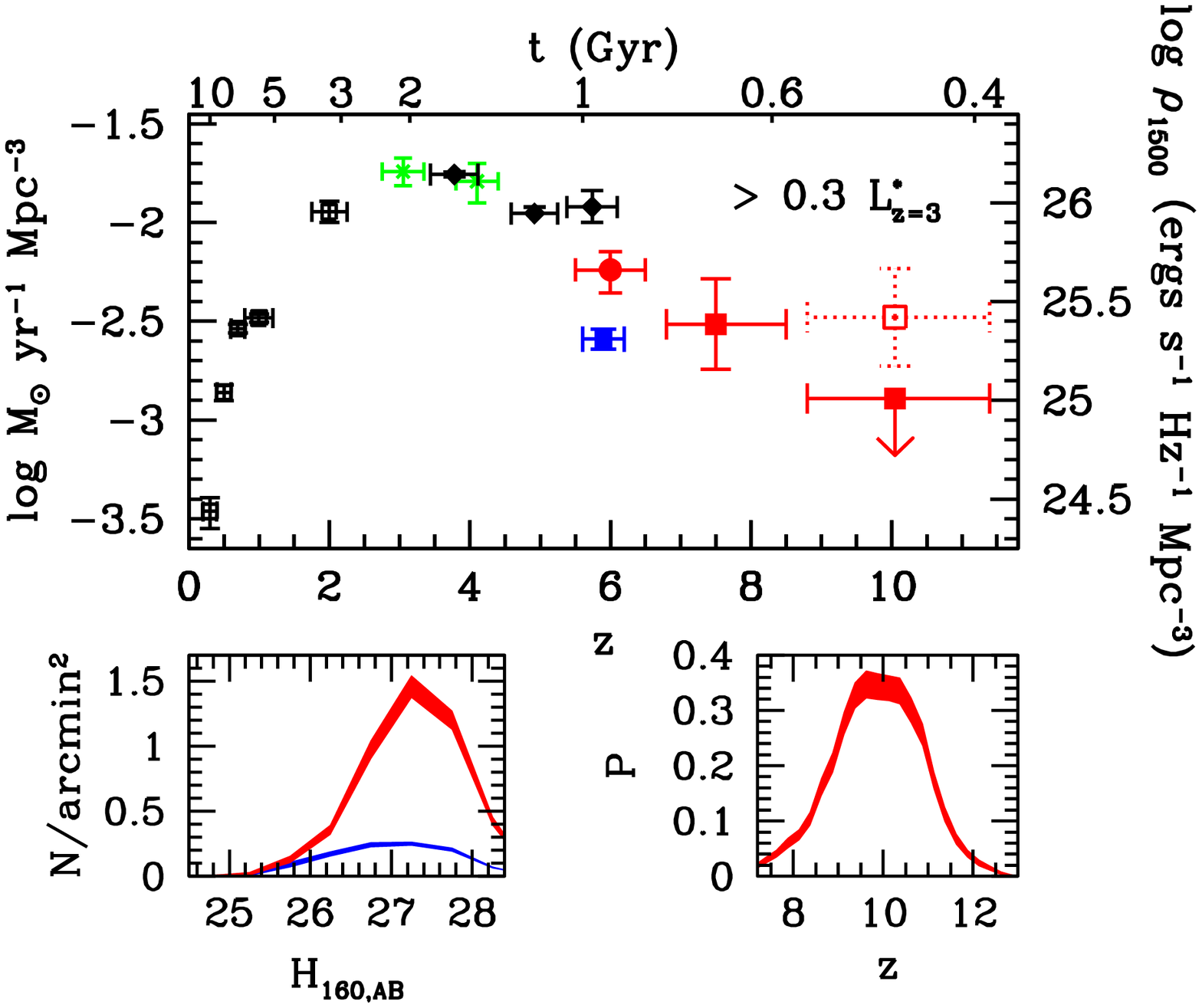}
\caption{The cosmic star formation rate density versus redshift with
no extinction correction.  The star formation rate density (integrated
down to 0.3 $L_{z=3} ^{*}$ -- the limit of our $z\sim10$ search) was
calculated from the luminosity density in the rest-frame UV continuum
($\sim1500\AA$: plotted on the right vertical axis) using canonical
assumptions (e.g., Madau et al.\ 1998) and a Salpeter IMF.  The open
red square at $z\sim10$ shows our result if 3 objects from this study
prove to be at $z\approx10$, while the large downward pointing arrow
shows our $1\sigma$ limits if none are (see text).  Included on this
plot are also estimates by Schiminovich et al.\ (2004) (open black
squares), Steidel et al.\ (1999) (green crosses), Giavalisco et al.\
(2004) (black diamonds), Bunker et al.\ (2004) (solid blue square),
Bouwens et al.\ (2004c) (solid red square), and Bouwens et al.\
(2005b) (solid red circle).  Consistent with past practice, the error
bars reflect poissonian uncertainties.  Large scale structure (cosmic
variance) would add an estimated $\pm$20\% for many of the lower
redshift points (e.g., Somerville et al.\ 2004), $\pm$50\% for the
$z\sim7.5$ point, and $\pm$19\% for the $z\sim10$ point.  Lower left:
The surface density of $J_{110}$-dropouts predicted from a
$(1+z)^{-1}$ size scaling of a $z\sim3.8$ $B$-dropout sample from
GOODS (B05) at the depths of the NICMOS parallels to the UDF (shaded
red regions).  The blue shaded regions show the equivalent surface
density for all our search fields.  The drop in the predicted counts
at fainter magnitudes arises from incompleteness.  Lower right: The
redshift distribution for objects satisfying our $J_{110}$-dropout
criterion in the above simulations (after distributing our ``cloned''
objects over the interval $z\sim7-13$ in proportion to the
cosmological volume element).  The primary conclusion to be drawn from
these results is that there is a significant deficit of bright objects
at $z\gtrsim6$ relative to that present at lower redshifts.}
\end{figure}

\section{Implications}

We have carried out a search for $z\approx10$ $J$-dropouts and found at
most 3 possible candidates.  Since $15.6\pm3.9$ and $4.8\pm2.2$
candidates are expected (Poisson errors) based on a $(1+z)^{-1}$
scaling of $z\sim3.8$ and $z\sim6$ samples, this suggests there has
been appreciable evolution at the bright end of the luminosity
function.  Since an $L_{z=3}^{*}$ galaxy at $z\sim10$ has a
$H_{160,AB}$-band magnitude of 26.7, our search
($H_{160,AB}\lesssim28$) tells us something about the galaxy
luminosities brightward of $0.3L_{z=3}^{*}$.  For simple Poissonian
statistics and assuming that all 3 sources are at $z\approx10$, the
current findings are inconsistent with no-evolution at the 99.98\% and
71\% confidence levels, respectively (equivalent to $3.8\sigma$ and
$1\sigma$).

Of course, cosmic variance is bound to be important for fields of this
size.  Assuming a selection window of width $\Delta z = 2.5$, a
$\Lambda$CDM power spectrum, and a bias of $7$--which corresponds to
the rough volume density of sources $\sim10^{-4}$ Mpc$^{-3}$ explored
by this probe (Mo \& White 1996)--we calculate $\sim$27\% RMS
variations field-to-field for single 0.8 arcmin$^2$ NIC3 pointings
(using a pencil beam window function).  Since our deepest two fields
provide the primary constraints and they are essentially independent,
the total number of $J$-dropouts found here is expected vary by
$\sim$19\% RMS relative to the cosmic average.  Thus, we expect
$\sim$$16\pm5$ and $\sim$$5\pm2$ dropouts, respectively, for simple
projections of our lower redshift samples, and so the present findings
are inconsistent with no-evolution at the 99.9\% and 68\% confidence
levels, respectively (equivalent to $3.3\sigma$ and $1.0\sigma$).

It is conventional to cast these findings in terms of the rest-frame
continuum $UV$ ($\sim1500\AA$) luminosity density.  We first estimate
the luminosity density under the assumption that the 3 candidates
(Figure 2) are at $z\approx10$.  The result is
$\rho_{UV}(z\sim10)/\rho_{UV}(z\sim3.8)=0.19_{-0.09}^{+0.13}$ (3 over
15.6 for Poissonian statistics) or
$\rho_{UV}(z\sim10)/\rho_{UV}(z\sim3.8)=0.19_{-0.10}^{+0.15}$
(including cosmic variance).  Taking the value of the luminosity
density at $z\sim3.8$ (Giavalisco et al.\ 2004) integrated down to
$0.3L_{z=3}^{*}$, we find a UV luminosity density
$\rho_{UV}(z\sim10)=2.5_{-1.2}^{+1.7} \times10^{25}$ erg s$^{-1}$
Hz$^{-1}$ (Poissonian statistics) or
$\rho_{UV}(z\sim10)=2.5_{-1.3}^{+2.0} \times10^{25}$ erg s$^{-1}$
Hz$^{-1}$ (including cosmic variance).  However, it is quite plausible
that none of these candidates are at $z\approx10$.  The expected high
contamination level combined with the now well established changes in
the LF between $z\sim6-7$ and $z\sim3$ suggests that it is more likely
that these sources are low redshift objects.  Assuming that none are
at $z\approx10$, the $1\sigma$ upper limit is:
$\rho_{UV}(z\sim10)/\rho_{UV}(z\sim3.8)<0.07$ (both for simple
Poissonian statistics and including cosmic variance).  In terms of the
luminosity density, this limit is
$\rho_{UV}(z\sim10)<0.9\times10^{25}$ erg s$^{-1}$ Hz$^{-1}$.
Adopting a Salpeter IMF and using canonical relations to convert this
into a star formation rate density (Madau et al.\ 1998), we can plot
the present determination (integrated down to $0.3L_{z=3}^{*}$)
against previous determinations at lower redshift (Figure 3).  The
observations now more clearly than before allow us to trace the number
of $UV$-bright systems over the interval $0<z<10$.  The space density
of high luminosity systems seems to peak at $z\sim2-4$ and decline
fairly rapidly to both higher and lower redshift.

As we conclude, it is somewhat sobering to realize that $\sim$740
orbits of deep NICMOS imaging went into this search and only 3
possible $z\approx10$ candidates were found (all of which may be at low
redshift), showing how difficult it is to map out the formation of
galaxies at these early times with current technology.  It is exciting
nevertheless to realize that in the future these surveys will be
executed much more efficiently.  For example, surpassing the current
NICMOS data set would require just $\sim$23 orbits with HST WFC3 and
$\sim1000-2000$ seconds with JWST.

\acknowledgements

We are grateful to Steve Beckwith and the entire STScI science team
for their foresight in taking deep NICMOS parallels to the UDF, Mark
Dickinson and STScI for making their NICMOS reductions available in
electronic form, Daniel Eisenstein, Bahram Mobasher, and Evan
Scannapieco for valuable discussions, Haojing Yan for an electronic
copy of a 2.5 Gyr ERO SED, and our referee for valuable comments which
substantially improved this manuscript.  This research was supported
under NASA grant HST-GO09803.05-A and NAG5-7697.

\newpage

\begin{deluxetable}{lrrrrrrrrr}
\tablewidth{0pt}
\tabletypesize{\footnotesize}
\tablecaption{$J_{110}$-dropout Search Fields.\label{tbl-1}}
\tablehead{
\colhead{} & \colhead{Area} & \multicolumn{2}{c}{$5\sigma$ limit} \\
\colhead{Field} & \colhead{($\sq'$)} & \colhead{$J_{110}$} & \colhead{$H_{160}$}}
\startdata
HDF-N Thompson & 0.8 & 27.8 & 28.1 \\
HDF-N Dickinson & 5.2 & 27.2 & 27.0 \\ 
HDF-S NIC-Par & 0.8 & 28.2 & 28.2 \\
UDF NIC-Par-1 & 1.3 & 28.6 & 28.5 \\
UDF NIC-Par-2 & 1.3 & 28.6 & 28.5 \\
UDF Thompson & 5.5 & 27.7 & 27.5 \\
\enddata
\end{deluxetable}

\begin{deluxetable}{lccccccccc}
\tablewidth{0pt}
\tabletypesize{\scriptsize}
\tablecaption{Red $(J_{110}-H_{160})_{AB}>1.8$ objects in our search fields.\tablenotemark{a}\label{tbl-1}}
\tablehead{
\colhead{Object ID} &
\colhead{R.A.} & \colhead{Decl.} &
\colhead{$H_{160}$} & \colhead{$J - H$} & \colhead{$V - H$} & \colhead{$z - H$} & \colhead{$H - K$} & \colhead{$H - m_{3.6\mu m}$} & \colhead{$r_{hl}$(\arcs)}}
\startdata
UDF-39188323 & 03:32:39.18 & -27:48:32.3 & 23.3$\pm$0.1 & 2.0 & 3.8 & 2.5 & 0.4 & 1.2 & 0.26 \\
UDF-42888094 & 03:32:42.88 & -27:48:09.5 & 24.8$\pm$0.1 & 2.7 & 4.3 & 3.3 & 1.0 & 2.9 & 0.25 \\
UDFNICPAR1-01191115 & 03:33:01.23 & -27:41:11.5 & 27.0$\pm$0.1 & 1.8 & 0.9 & $>$-0.2 & --- & --- & 0.24 \\
UDFNICPAR1-04151142 & 03:33:04.18 & -27:41:14.1 & 28.2$\pm$0.2 & 1.8 & $>$-0.0 & $>$-1.3 & --- & --- & 0.18 \\
UDFNICPAR1-05761077 & 03:33:05.80 & -27:41:07.6 & 27.8$\pm$0.2 & $>$2.0 & $>$0.3 & $>$-1.0 & --- & --- & 0.21 \\
UDFNICPAR2-07352493 & 03:33:07.35 & -27:52:49.3 & 27.9$\pm$0.2 & 1.8 & 0.6 & $>$-0.7 & --- & --- & 0.14 \\
UDFNICPAR2-09593048 & 03:33:09.59 & -27:53:04.8 & 27.3$\pm$0.2 & 1.9 & $>$0.9 & $>$-0.4 & --- & --- & 0.19 \\
HDFN-46242594 & 12:36:46.24 & 62:12:59.4 & 26.5$\pm$0.1 & $>$2.1 & 1.1 & --- & $<$1.7 & --- & 0.17 \\
HDFN-51763567 & 12:36:51.76 & 62:13:56.7 & 25.0$\pm$0.1 & 2.0 & 2.1 & --- & 0.6 & --- & 0.31 \\
HDFN-56253227 & 12:36:56.25 & 62:13:22.7 & 25.2$\pm$0.1 & $>$3.2 & $>$4.2 & --- & 1.5 & 2.0 & 0.20 \\
HDFSPAR-48278437 & 22:32:48.27 & -60:38:43.7 & 26.0$\pm$0.1 & $>$2.5 & --- & --- & 1.5 & --- & 0.32 \\
\enddata
\tablenotetext{a}{All astrometry uses the J2000 equinox.  Magnitudes
are in the AB system.  All limits are 2$\sigma$ except the limits on
the $J_{110}-H_{160}$ color which are only 1$\sigma$.  Total
magnitudes and colors are derived in Kron apertures of different sizes
(B05).  Object IDs are the last four significant figures given in this
table for the Right Ascension and Declination.}
\end{deluxetable}

\end{document}